\documentclass[vecphys]{svmult}

\usepackage{makeidx}         
\usepackage{graphicx}        
\usepackage{multicol}        
\usepackage[bottom]{footmisc}

\makeindex             

\begin{document}

\title*{Science with the World Space Observatory - Ultraviolet}
\author{Ana I. G\'omez de Castro\inst{1} 
\and Isabella Pagano \inst{2} \and Mikhail Sachkov
\inst{3} \and Alain Lecavelier \inst{4} \and Giampaolo Piotto \inst{5} 
\and Rosa Gonz\'alez \inst{6} \and Boris Shustov \inst{3} }
\authorrunning{G\'omez de Castro et al.}
\institute{Fac. de CC. Matem\'aticas, Universidad Complutense de
Madrid, Plaza de Ciencias 3, E-28040 Madrid, Spain
\texttt{aig@mat.ucm.es}  \and INAF - Osservatorio Astrofisico di Catania, v. S. Sofia 78, 95123 Catania, Italy \texttt{isabella.pagano@oact.inaf.it}
\and Institute of Astronomy of the Russian Academy of Science, 48 Pyatnitskaya St.,
119017 Moscow, Russia \texttt{sachkov@inasan.ru}
\and Institute d'Astrophysique de Paris, 98bis, bd Arago - 75014 Paris  FRANCE 
\texttt{lecaveli@iap.fr} 
\and Dipartimento di Astronomia, Universit\`a di Padova, Vicolo dell'Osservatorio, 3
35122 Padova, Italy \texttt{giampaolo.piotto@unipd.it}
\and Instituto de Astrofisica de Andalucia, Camino Bajo de Huetor 50,
E-18008 Granada, Spain \texttt{rosa@iaa.es}
}
%
%
\maketitle

\begin{abstract}
The World Space Observatory-Ultraviolet (WSO-UV) will provide
access to the UV range during the next decade. The instrumentation
on board will allow to carry out high resolution imaging, high
sensitivity imaging, high resolution (R$\sim 55000$) spectroscopy
and low resolution (R$\sim 2500$) long slit spectroscopy. In this
contribution, we briefly outline some of the key science issues
that WSO-UV will address during its lifetime.
Among them, of special interest are: the study of galaxy formation and the
intergalactic medium; the astronomical engines; the Milky Way formation and evolution, and the formation of the Solar System and the atmospheres of extrasolar planets.

\end{abstract}

\section{Introduction}
\label{sect:intro}  

Launched at the beginning of the next decade, the World Space Observatory 
- Ultraviolet (WSO-UV) will fly for 5 years with a probable extension of other 5 years.
The observatory will be in a circular geosynchronous orbit with
35\,800~km radius, with 51.8 degrees inclination;
Earth occultation will be small and the orbital period
will allow to time-track targets and to have rapid access to
targets of opportunity.

The WSO-UV will be equipped with multipurpose instrumentation to carry
out:

\begin{itemize}
\item High resolution spectroscopy (R$\sim 55000$) of point sources
in the range 102-320~nm by means of two high resolution echelle
set-ups - HIgh Resolution Double Echelle
Spectrograph (HIRDES). The
sensitivity of this instrument is 10 times better that of the
Space Telescope Imaging Spectrograph (STIS) on a similar
configuration.

\item Long-slit low resolution (R$\sim 1500 - 2500$) spectroscopy in the
102-320~nm range.

\item High resolution (0.03 arcsec pixel$^{-1}$) imaging in the 150-280~nm range with
a field of view of 1$\times$1 arcmin$^{2}$.

\item High resolution (0.07 arcsec pixel$^{-1}$) imaging in the 200-700~nm range with
a field of view of 4.7$\times$4.7 arcmin$^{2}$. This camera will be in average a
factor of 3 more effective {\sl sensitive} than the Advanced Camera System (ACS)
and the Wide Field Camera 3 on the Hubble Space Telescope (HST).

\item A large field (6.6$\times$6.6 arcmin$^{2}$) far ultraviolet (115-190~nm) camera
with spatial resolution 0.2 arcsec pixel$^{-1}$. This will allow Lyman-$\alpha$, C~IV
or molecular hydrogen mapping in extended, faint, astrophysical
targets.

\end{itemize}
General information on the WSO-UV mission are given by Sachkov et al. 
(this proceedings), while detailed information on the focal plane 
instruments can be found in Pagano et al. (this proceedings).

The Ultraviolet (UV) is a fundamental
energy domain since it grants high sensitivity access to the study
of both atomic plasmas at temperatures in the 3,000-300,000~K
range and  molecular plasmas illuminated by UV radiation since the
electronic transitions of the most abundant molecules in the
Universe (H$_{2}$, CO, OH, CS, CO$_{2}^+$, C$_{2}$...) are in this
range; the UV radiation field itself is a powerful astrochemical
and photoionizing agent. For this reason, most of the areas
of astrophysics will benefit from the availability of a
world-class observatory as WSO-UV during the next decade.
A detailed  accounting of the needs for  UV instrumentation in the
many fields of astrophysical research can be found in the special
issue entitled {\it Fundamental problems in astrophysics:
requirements for future UV observatories} (G\'omez de Castro \&
Wamsteker, eds., 2006) that summarizes  the thoughts and work of
the {\it Network of UV Astronomy}
(NUVA\footnote{http:www.ucm.es/info/nuva/}) and the UV community
at-large.

The WSO-UV will work as a Space Observatory with a core program,
guaranteed time for the project partners and time open to excellent
scientific projects from the world-wide community. In this contribution
we outline some of the key issues in the WSO-UV Observatory program:

\begin{enumerate}

\item Galaxy formation: determination of the diffuse baryonic content
in the Universe and its chemical evolution.

\item The physics of accretion and outflow: the astronomical engines.

\item The Milky Way formation and evolution

\item Extrasolar planetary atmospheres and astrochemistry in the
presence of strong UV radiation fields.

\end{enumerate}

\section{Determination of the diffuse baryonic content
in the Universe and its chemical evolution}

\label{sect:WIGM}

The composition of the Universe (baryonic matter, dark matter and
dark energy densities) is derived from observations related to the
first 3~Gyr the Universe (Cosmic Background, high z quasars and
ultraluminous galaxies at $z>3$). To understand the physical transition  from this
early epoch to our current Universe (14 Gyr) it is necessary to
get information on the recycling of the gas (InterGalactic Medium,
IGM) and the evolution of the stellar population and the radiation
field  over   the redshift interval $0<z<2$ that spans a 80\% of
the lifetime of the Universe. Though theoretical modelling on
structure and galaxy formation is progressing rapidly there are
no data to contrast with since the IGM is poorly studied.
Fundamental parameters such as the galaxy formation efficiency and
its variation across the Universe cannot be {\it measured} since
data about the gas component are, at the very least, scarce.
Fundamental questions such as, which physical forces trigger
galaxy formation or what controls the galaxy mass function and the
lower limits to galactic masses cannot be addressed without a good
knowledge of the IGM properties.

UV spectroscopy and imaging are  the most sensitive technique to obtain direct
information in the $z<2$ Universe on:

\subsection {The baryonic content of the Warm-Hot IGM}

Independent of the different models for the early Universe the
major baryonic component of the Universe at~$z<3$ must be
associated with the IGM. The Warm-Hot component of the IGM (WHIM),
at temperatures $T=10^5-10^7$K, most likely accounts for a $\sim
30$\% of the baryons to the cosmological mass density in the local
Universe (see {\it e.g.} the cosmological simulations by Cen \&
Ostriker 1999). 
Broad HI-Ly$\alpha$ absorption is one of the most promissing 
techniques to map the distribution of the WHIM.

\subsection{The Damped Ly-$\alpha$ (DLA) systems }

DLA systems trace column densities about 10$^{20}$~cm$^{-3}$, {\it
e.g.} dense clumps of gas likely associated with galaxy formation
sites. Numerical simulations show that these clumps arise
naturally in the hierarchical theory of galaxy formation (Katz et
al 1996)  and thus become the best tracers to the characteristic
scales of structure formation between the early  epochs and the
local Universe. DLAs trace the location of the gravitational
potential wells of all the matter, henceforth representing another
tracer of dark matter.

\subsection{He\,II reionization}

From theoretical modeling of the IGM we know that after the HI
reionization, the IGM cools by expansion, is reheated  by the
delayed HeII reionization at z=3, and continues to cool with
decreasing redshift. Observations of the HeII$\lambda$304\AA\
forest over the redshift range $2.1<z<2.9$ will test this model in
the most direct way (Reimers et al 2005).

\subsection{The role of starbursts in the evolution of the IGM}
Starbursts are systems with very high star formation rate per unit area. 
They are the preferred place where massive stars form; the main source 
of thermal and mechanical heating in the interstellar medium, and the 
factory where the heavy elements form. Thus, starbursts play an important 
role in the origin and evolution of galaxies. The similarities between 
the physical properties of local starbursts and high-$z$ star-forming galaxies, 
highlight the cosmological relevance of starbursts. Moreover, 
nearby starbursts are laboratories where to study violent star formation 
processes and their interaction with the interstellar and intergalactic media, 
in detail and deeply. Outflows in cold, warm and coronal phases leave 
their imprints on the UV interstellar lines. Outflows of a few hundred 
km~s$^{-1}$ are ubiquitous phenomena in starbursts. These metal-rich 
outflows and the ionizing radiation can travel to the halo of galaxies and 
reach the intergalactic medium. Observations with the WSO-UV will allow
to study the stellar content and the high mass IMF of the starburst 
and to determine the chemical composition of the associated galactic 
winds.

\subsection{Galaxies formation}
How the present-day Hubble sequence of galaxies was assembled, 
as traced by the buildup of their mass, gas, stars, metals, and magnetic fields?
Understanding the formation and evolution of galaxies is driving most of the 
current efforts in observational cosmology.  Galaxies are thought to form inside 
the dark matter (DM) halos produced by gravitational instability from the seed 
initial fluctuations.  The evolution of the DM component is successfully predicted 
from first principles, because it evolves under the sole action of gravity.  However, 
the predictive power of theory drops dramatically when it comes to the luminous, 
baryonic component of the universe.  Complex physical processes come into play, 
and it is no surprise that the attempts at modelling galaxies have so far attained 
limited success.
A number of different investigations with WSO-UV imagers can provide fundamental 
inputs to the problem of galaxy formation and evolution. 
For instance, high resolution 
and deep images in wide bands at 150 nm, 220 nm, and 300nm imaging of GOODS/UDF Deep 
Fields (Dickinson et al. 2003),  allow to:
\begin{enumerate}
\item Search for Lyman Break Galaxies at relatively low z. While the search for LBG 
galaxies is a very well established technique at z$\ge$2-3, the interval around 
z$\sim$1--1.5 is poorly explored due to the lack of deep and good angular resolution 
observations.
\item Study of the UV luminosity function: the UV (rest frame 150 nm) luminosity 
function is currently poorly constrained at the faint end. The 300 nm observations 
will fill the gap in particular at z$\sim$1-1.5, where the global cosmic star formation 
rate stars to decline.
\end{enumerate}

\section{Astronomical engines}

Astronomical engines (stars, black holes, etc...) can accelerate
large masses to velocities close to the speed of light  or
generate sudden ejections of mass as observed in Supernova
explosions. They are also able to produce significantly milder
winds, as seen in the Sun, or to eject gas shells induced by
pressure pulsations in the stellar atmosphere. All of these
phenomena transform energy of various forms (gravitational,
thermal, radiative, magnetic) into mechanical energy to produce
outflows in conditions very different to those tested in Earth
laboratories.  The least conventional engines are those generating highly
collimated bipolar outflows and jets. These are thought to be
driven by a combination of gravitational energy, differential
rotation and magnetic fields. They are among the most exciting
objects in nature; however, their underlying physics is poorly
known. This physical regime affects all of the many scales of
Astrophysics; it determines the luminosity of the Active Galactic 
Nucleii and the re-ionization of the Universe at $z\simeq 3$. It also determines
the properties of planetary systems, which are just angular
momentum reservoirs left over when the engine is turned off in
pre-main sequence stars. 

WSO-UV will provide key inputs to answer the basic open questions
concerning this physics:
\begin{enumerate}

\item What controls the efficiency of accreting objects as gravitational
engines?, is the magnetic field needed to guarantee that outflows
are fast?, what are the relevant timescales for mass ejection?

\item How does the accretion flow proceed from the disk to the
source of the gravitational field in the presence of moderate
magnetic fields?, which fraction of the gravitational energy lost
in this process is deposited on the stellar surface?, which
fraction is lost in amplification/dissipation of magnetic flux?

\item Which is the role played by radiation pressure in this
whole process?

\item What role do disk instabilities play in the whole accretion/outflow
process?, which are the key mechanisms driving these
instabilities?

\end{enumerate}

\noindent
Some few examples are:
\begin{enumerate}
\item High resolution UV spectroscopy will allow to determine 
the structure of the accretion flow on
magnetic cataclismic variables  and on T Tauri stars and to
measure the physical conditions and clumpyness of the outflows. 
It will also allow to study the source of energy
that powers  the extended dense ($\geq 10^{10}$cm$^{-3}$)
and hot (T$_e \geq 60,000$~K) envelopes that have been
detected  around T Tauri stars (G\'omez de Castro \& Verdugo, 2003). 
The luminosities of these envelopes are about 
0.2 L$_{\odot}$.

\item Low resolution spectroscopy will allow to measure
the general physical conditions and 
metalicities of the Broad
Lines Emission Region in Active Galactic Nucleii. Reverberation
mapping will allow to study the kinematics and the mass of the
central supermassive black holes. Also the atmospheres
of the hot accretion disks in cataclysmic variable stars
will be studied and the role of disk instabilities in
triggering the outbursts.

\item The high sensitivity FUV camera will allow to
detect hot jets through their Ly-$\alpha$ and
to resolve the thermal structure of the jets and
the regions shocked by them. It will also allow to
survey star forming regions to detect planetary-mass
objects in regions like $\sigma$-Orionis and to study
the magnetic activity and accretion processes in
these free-floating planetary-like objects which
are at the low mass end of the molecular clouds 
fragmentation scales.

\end{enumerate}

\section{How did our Galaxy form?}

What is the detailed history of the formation and evolution of our own 
Galaxy, and what lessons it hold for the formation and evolution of galaxies generally.
Globular clusters have traditionally been considered as good tracers 
of the process that led to the formation of their host galaxy. Absolute ages 
set a constraint on the epoch of the galaxy formation, with important cosmological 
consequences. Relative ages provide detailed information on the formation process of 
the host galaxy. For the absolute age determination, accurate distances, metallicity, 
and reddening are mandatory. As for the distance, while waiting for GAIA results, 
there is a simple, geometric method to get accurate distances of GCs: the comparison 
of the dispersion of internal proper motions, an angular quantity, with that of radial 
velocities, a linear quantity gives a distance. Radial velocities for thousands of stars 
in GCs, with an accuracy of a few hundred km/s are now easily attainable with multifiber 
facilities at the 8--10m class telescopes. As for the proper motions, already King et al. 
(1998) pioneeristically showed that the WFPC2 on board of HST allows astrometric position 
measurements with a precision of the order of a few milliarcseconds (mas) on a single image. 
With WSO-UV FCU imagers we expect to reach accuracies of $<$1\,mas/frame in the UVO channel 
(200-700 nm). This accuracy, taking into account the possibility of using the archive HST 
data, i.e. the fact that we have a very long temporal baseline (20--30 years), allows to 
obtain relative proper motions measurements with accuracies of $<$10 microarcsec/yr.
With the high accuracy proper motion and radial velocity measurement, the distance 
uncertainty mainly depends on the sampling error, which goes with the (2n)$^{1/2}$, 
where n is the number of measured stars (typically a few thousands).
Using background nucleated galaxies (or, even better, background point like sources 
like QSO) it is also possible to have absolute proper motions with an accuracy which 
is mainly related to the ability to measure the position of the reference galaxies. 
This accuracy can be estimated to be of the order of a 30 microarcsec (better in the 
case that the reference objects are background QSOs).
With these absolute proper motions it is possible to:
\begin{enumerate}
\item use the 3-D kinematics of GCs as a probe of the Galactic potential;
\item use the 3-D kinematics, coupled with information on the GC ages and metal 
content, to detect Galactic streams that are important information to understand 
the role of the accretion from small size satellites in the assembly of our Galaxy.
\end{enumerate}

\section{Young planetary disks and extrasolar planetary atmospheres}

T~Tauri stars (TTSs), solar-like pre-main sequence stars, are
unique to study the environment (radiation, high energy particles,
dynamical processes) in which planetary systems, like ours, grow.
Notice that recent theories propose that the inner, Earth-like,
planets begin to build-up some 10$^6$yr after the star begin to
form and, at this stage, the accretion-based engine is still
operating. The radiation produced by the engine ought to have an
important effect on the inner disk evolution and the evaporation
of the primary atmospheres of the planets-embryos through
photoionization and photochemical reactions (Watson et al 1981,
Lecavelier des Etangs et al 2004). Thus, UV spectroscopy will
allow the study of the interactions between the stellar UV field
and  the young planetary disks and detect the molecular component
in very diffuse disks (Jolly et al 1998).
Very recent chemical models are showing that the penetration of UV
photons coming from the central engine in a dusty disk could
produce an important change in the chemical composition of the gas
allowing the growth of large organic molecules. In this context,
UV photons at $\lambda > 1500$~\AA\ photodissociating organic
molecules could play a key role in the chemistry of the inner
regions of the proto-planetary disk, while those photodissociating
H$_2$ and CO would control the chemistry of the external layers of
the disk directly exposed to the radiation from the star. The
radiation field can produce a rich photochemistry on timescales
shorter than the dynamical evolution time scales, leading to the
formation of large carbon-rich molecules such as C$_n$H$_2$,
HC$_{(2n+1)}$N, and C$_n$. Reactions between these species and H
and H$_2$ may maintain their high abundances in spite of the
strong radiation field emerging from the central star.

UV spectroscopy will also allow to study the properties of the
atmospheres of extrasolar planets (high resolution ultraviolet
spectroscopy has been the technique that has allowed
detecting/studying for the first time the atmosphere of an
extrasolar planet: HD~209458b).  Observations of exoplanets and the detailed
characterization of their atmospheres will help us understand
better the physical processes at work in the building of a planet
and its atmosphere, and in the further evolution of such a system.

The WSO-UV imager will offer the chance to search for and study 
the ultraviolet auroral emission generated by extrasolar giant planets: 
i) it allows a direct detection of a planet instead of the indirect 
methods employed to date (radial velocity and pulsar timing); ii) the 
presence of an auroral signature is strictly linked to the presence of 
a planetary magnetic field and the evidence of this effect can not be 
observed with any other detection method; iii) ultraviolet auroral could 
contribute to characterize the near space environment around planet, 
so that their study could provide information about both basic atmospheric 
composition and the energies of the impacting particles. In general, 
UV wavelengths provide observational advantages compared to the optical. 
In fact, higher contrast ratios can be achieved in UV and the UV diffraction 
limit allows planets to be detected at smaller angular separations to 
their host stars.

In the coming decade, several ground and space-based observing
programs will lead to the discovery of an extremely large number
of exoplanets, in particular, near-future  space missions
including Corot, Kepler or GAIA will discover large numbers of
exoplanets transiting their parent stars. To acquire a revealing
picture of these new worlds, we need to characterize the planetary
atmospheres of a large sample of these exoplanets. The observation
of UV and optical absorptions occurring when an exoplanet transits
its parent star are a very powerful diagnostic technique because
of the strong absorption of stellar UV photons by the ozone
molecule in the planetary atmosphere (see Ehenreich et al 2005).

In this manner, we expect to be able to address some important
questions as: how do properties such as effective temperature,
stellar type, high-energy particle environment, and metallicity of
the central star alter the evolution of its planetary system? What
effects do a planet's orbital parameters (orbital distance and
eccentricity) have on its size, mass and potential migration
during the formation process?(see G\'omez de Castro et al 2006 for
more details).

\section{Conclusions}

The scientific plans for WSO-UV are very ambitious, and span all 
of the astronomical research branches. WSO-UV will be operating in 
the second decade of this century, and it will be a fundamental tool 
for the development of astronomical knowledge, fully integrated with 
the many other space and ground-based observatories (including the new 
generation ELT telescopes) operating in the same temporal interval. The 
space missions operating in the WSO-UV era will provide observational 
data both at shorter wavelengths (e.g. Symbol-X, possibly the extended 
XMM and Chandra missions, etc.) and longer wavelengths (e.g. GAIA, JWST, 
Herschel etc.) than those covered by an UV optimized mission like WSO-UV. 
WSO-UV observations are a necessary, fundamental complement 
of the data set collected by presently operating and planned space 
and ground based observatories.

\subsection{Acknowledgments}
Authors are grateful to the members of the Network for Ultraviolet
Astronomy (NUVA) and to the other WSO--UV team members.
AIGdC acknowledges the support by the
Ministry of Science and Education through grant ESP2006-27265-E.
IPa and GPi thank colleagues of the Italian WSO--UV Science Team for their valuable 
contribution, in a special way M.Nonino, and S.Marchi. The participation in the 
WSO-UV project in Italy is funded by Italian Space Agency under contract 
ASI/INAF No. I/085/06/0.


\end{document}